\documentstyle[12pt]{article}
\def \be{\begin{equation}}
\def \ee{\end{equation}}                
\def \ba{\begin{array}{l}}
\def \ea{\end{array}}
\def \bq{\begin{eqnarray}}
\def \eq{\end{eqnarray}}
\def \nn{\nonumber\\}
\def \lb{\label}
\def \ln{{\rm ln}}

\def \fr{\frac}

\def \ol{\overline}

\def \la{\langle}
\def \ra{\rangle}
\def \[{\left[}
\def \]{\right]}
\def \({\left(}
\def \){\right)}
\def \2{\frac{1}{2}}
\def \4{\frac{1}{4}}
\def \a{\alpha}
\def \b{\beta}

\def \d{\delta}

\def \e{\epsilon}

\def \T{T_{c}}

\newcommand{\sectio}[1]{\section{#1}\setcounter{equation}{0}}

\begin{document}

\frenchspacing
\setlength{\parskip}{2mm}
.


\vspace{10mm}

\begin{center}

{\Large \bf Mean Field Model of a Glass}
\vskip .2in
Victor Dotsenko\footnote{On leave from 
Landau Institute for Theoretical Physics, Moscow}
\vskip .1in
Laboratoire de Physique  Theorique des Liquides, \\
UMR 7600, Universite Paris VI, \\
4 place Jussieu, 75252 Paris Cedex 05,
France\\

\end{center}

\vskip .3in

\begin{abstract}
In this paper we propose a simple mean-field "toy" model for 
the liquid-glass phase transition. This is the system 
of $N$ point-like particles confined in a finite volume 
of a $D$-dimensional space interacting via infinite-range 
oscillating potential. In the framework of the replica 
approach it is shown that such a system exhibits the phase 
transition between the high-temperature liquid phase and the
low-temperature glass phase. This phase transition is described 
in terms of the standard one-step replica symmetry breaking 
scheme. 
\end{abstract}

\newpage

\sectio{Introduction}

The problem of the liquid-glass phase transitions attracts permanent
interest during last decades (for reviews see e.g. \cite{glass}).
In the recent years
in additional to the traditional experimental and phenomenological
investigations a notable progress has been achieved in 
a first principle statistical mechanical study of the glass phase
(see \cite{MP} and references therein).
Leaving apart a wide scope of non-equilibrium properties of glasses, in
a pure statistical mechanical approach one is aiming to investigate
the thermodynamical properties of the $N$-particle system
with the two body interparticle interactions 
described by the Hamiltonian

\be
\lb{g1}
 H[{\bf x}_{i}] = \sum_{i,j=1}^{N} U({\bf x}_{i} - {\bf x}_{j}) \; ,
\ee
where ${\bf x}_{i}$ is the vector in a $D$-dimensional space 
which points the position of the $i$-th particle, and $U({\bf x})$
is the interparticle potential (one, of course, can consider more
general systems with two or more sorts of particles and different 
potentials for different particles). In a realistic systems the potential
$U({\bf x})$ must be attractive (and sufficiently quickly decaying) 
at large distances and strongly repulsive at short distances. 

In a very simplified form the scheme of calculations is supposed
to look as follows. 
Instead of the plain (and hopeless) integration
over all positions of the particles in the partition function

\be
\lb{g2}
Z = \[\prod_{i=1}^{N}\int d^{D} {\bf x}_{i}\]
        \exp\(-\b H[{\bf x}_{i}]\)
\ee
in the systematic approach, first of all, one should find (or rather guess) 
the space structure and the energy of the (zero-temperature) 
ground state configuration of the Hamiltonian,
and then, using one or another approximation, 
integrate over the fluctuations around this state. 
Depending on the value of the temperature one eventually finds 
that either the proposed non-trivial ground state is stable
with respect to the fluctuations (at low enough temperatures), or
(at sufficiently high temperatures) the fluctuations destroy 
proposed ground state and the thermodynamic state of the system is
a liquid where all the particles are delocalized.

Indeed, the above scheme works rather well if the low temperature
ground state of the system is ordered. In this case we are dealing with
the crystal which is characterized by one or another global
symmetry breaking, and which can be sufficiently easily described 
analytically. However, in the study of the disordered
glass state, the situation becomes much more complicated,
because in this case the low-temperature solid state is characterized 
by the {\it random} positions of the particles, and it is impossible to
define what kind of the global symmetry is broken. All that resembles
a lot the problem we are facing in spin-glasses, where the spins are
getting frozen in a random state which can not be characterized
by any apparent global symmetry breaking. The only but quite essential
difference is that, unlike spin-glasses, here we do not have
{\it quenched disorder} installed in the initial Hamiltonian.
Nevertheless, the ideas borrowed from the spin-glass theory,
and in particular the use of the replica technique (which, as we
realize now, has much deeper meaning than just a technical trick
to pass over the averaging of the logarithm of the partition function)
turned out to be quite fruitful also for structural glasses,
as was shown in the series of papers by Mezard and Parisi \cite{MP}.

To demonstrate the effect of a symmetry breaking in ordered magnetic systems 
one can introduce a conjugated field coupled to the order parameter, 
which at the end (after taking the thermodynamic limit) is set to zero.
In spin-glasses the same effect can be
achieved by introducing several weakly coupled copies (replicas)
of the original system. In a similar way, to demonstrate the effect 
of the freezing into a random glass state in the system of particles
described by the Hamiltonian, eq.(\ref{g1}), let us
introduce {\it two identical copies} of the same system described 
the following Hamiltonian

\be
\lb{g4}
H = \sum_{i,j}^{N} U({\bf x}_{i} - {\bf x}_{j}) 
  + \sum_{i,j}^{N} U({\bf y}_{i} - {\bf y}_{j}) + 
 \e \sum_{i}^{N} W({\bf x}_{i} - {\bf y}_{i}) \; ,
\ee
which contains a weak (controlled by the parameter $\e$) attractive 
potential $W$ between particles $\bf{x}$ and $\bf{y}$
of the two systems. After taking the thermodynamic limit the parameter 
$\e$ must be set to zero. This trick can result in 
two types of situations: 

(1) After taking the limit $\e \to 0$ the 
particles of the two systems becomes completely independent 
(uncorrelated). This would indicate that the particles 
do not have a "memory" of their spatial positions so that
they are free to move (non localized in space) and the original 
system is in the (high-temperature) liquid phase.

(2) Even after taking the limit 
$\e \to 0$ the positions of the particles of the two systems remain 
correlated. This would indicate that the particles become
localized in space so that the original 
system is in the (low-temperature) solid state.
The order parameter describing this phase 
can be defined e.g. in terms of the correlation function between 
particles of the two copies of the system.

In fact, like in spin glasses, to obtain more detailed information about 
this type of phase transition, instead of the two copies it is more convenient 
to introduce a general  $n$ replicas of the original system (see Section 2). 
It should be also noted that in real calculations the introduction of 
the supplementary attractive potential between replicas is actually 
not necessary. It is well known that e.g. in the case of
the paramagnetic-ferromagnetic phase transition, instead of introducing
a conjugated field, it is sufficient just {\it to suppose} the possibility
of the global symmetry breaking to prove its existence afterwords.
In a similar way, here we are also going {\it to admit} the possibility 
of the effective space correlations among particles of different 
(originally non-coupled) replicas,
while the validity of this assumption can be checked aposteriori.

Although the above idea is very simple, actual calculations
for realistic models turn out to be quite sophisticated 
(see e.g. \cite{MP}). What is lacking in this field is the analog of the 
mean-field models of spin-glasses, like the SK-model \cite{SK} or REM 
\cite{rem}, which despite of their non-physical nature, turned out
to be very powerful tool for understanding the nature of the
spin-glass state.
The aim of this paper is to present a very simple {\it mean-field}
model of a glass, which, despite of its completely non-physical
structure, exhibits non-trivial liquid-glass phase transition
(Section 3). Moreover, the nature of this phase transition turns
out to be of the so-called one-step replica symmetry 
breaking type (the one, which, in particular, takes place in REM),
which is characterized by a finite jump of the order parameter
(like at the first-order phase transitions) and continuous free
energy function (as it should be at the second-order phase transition).

\sectio{Replica Calculations}

The general scheme of the replica calculations (which we will follow
in the next section) can be described as follows. First, we take $n$ 
non-coupled copies of 
the original system. The corresponding replica partition function is

\be
\lb{g7}
Z_{n} = \[\prod_{i=1}^{N}\prod_{a=1}^{n} \int d^{D} {\bf x}_{i}^{a}\]
        \exp\(-\b \sum_{a=1}^{n} H[{\bf x}_{i}^{a}]\)  \; ,
\ee
where the parameter $n$ should be set to $n=1$ et the end.
Next, following the standard scheme of the {\it one-step replica symmetry breaking}
\cite{mpv} we divide $n$ replicas into $n/m$ groups each consisting of $m$
replicas. We will suppose that particles belonging to different groups
of replicas are non-correlated, while those of the same group could become
effectively coupled. In this case the partition function, eq.(\ref{g7}),
reduces to

\be
\lb{g8}
Z_{n} = \[ Z_{m}\]^{\fr{n}{m}} \; ,
\ee
and the density of the free energy of the system is then

\be
\lb{g9}
f = -\fr{1}{\b N} \ln\(Z_{n}\) = -\fr{n}{\b m N} \ln\(Z_{m}\)
\ee
After taking the limit $n \to 1$ we get:

\be
\lb{g10}
f(m,\b) = -\fr{1}{\b m N} \ln\(Z_{m}\)
\ee
where $m$ becomes now the continuous variational parameter bounded by the 
condition $m \leq 1$ (this constraint comes from the original bound
$m \leq n$ in the limit $n \to 1$). 

The assumption that the positions ${\bf x}_{i}^{a}$ of particles in 
different replicas are correlated can be explicitly represented as follows:

\be
\lb{g11}
{\bf x}_{i}^{a} = {\bf x}_{i} + {\bf u}_{i}^{a} \; ,
\ee
where ${\bf x}_{i}$ play the role of the center of mass of the
replica "molecules", and ${\bf u}_{i}^{a}$ are the deviations of the
particles from the center of mass.
Of course, these deviations are bounded by the condition

\be
\lb{g12}
\sum_{a=1}^{m} {\bf u}_{i}^{a} = 0
\ee
In this way the replica partition function takes the following form:

\be
\lb{g13}
Z_{m} = \[\prod_{i=1}^{N}\int d^{D} {\bf x}_{i}\]
        \[\prod_{i=1}^{N}\prod_{a=1}^{m} \int d^{D} {\bf u}_{i}^{a}\]
        \[\prod_{i=1}^{N} m^{D} \d\(\sum_{a=1}^{m} {\bf u}_{i}^{a}\)\]
        \exp\(-\b \sum_{a=1}^{m} H[{\bf x}_{i} +{\bf u}_{i}^{a} ]\)
\ee
We see that now the problem becomes similar to that of statistical systems
with quenched disorder in replica representation: according to eq.(\ref{g13}), 
${\bf x}_{i}$'s play the role of disorder parameters, while ${\bf u}_{i}^{a}$
are the dynamical variables. As usual, after averaging over the disorder parameters
$\{{\bf x}_{i}\}$ we will get the partition function $Z_{m}$ represented
in terms of a new replica Hamiltonian $H_{m}[{\bf u}_{i}^{a} ]$:

\be
\lb{g14}
Z_{m} = \[\prod_{i=1}^{N}\prod_{a=1}^{n} \int d^{D} {\bf u}_{i}^{a}\]
        \exp\(-\b H_{m}[{\bf u}_{i}^{a} ]\) \; ,
\ee
where the variables ${\bf u}_{i}^{a}$ with different replicas
could become coupled.

In the final step of calculations will be looking for the extremum
of the free energy, eq.(\ref{g10}), with respect to the continuous
parameter $m$, bounded by the condition $m \leq 1$. As usual in the
replica calculations,  after the parameter $m$ 
(which originally was introduced to describe the number of replicas) 
is taken to be smaller than one, the minimum of the free energy
becomes a maximum. In the present case due to the constraint, eq.(\ref{g12}),
the number of independent replica variables ${\bf u}_{i}^{a}$
is equal to $(m-1)$. After taking the parameter $m$ to be smaller than one
this number formally becomes negative which effectively changes the minimum
of the free energy into the maximum (this situation is quite similar to that
of the standard replica calculations in disordered systems \cite{mpv})

The maximum of the free energy, eq.(\ref{g10}), is defined by the
saddle-point equation 

\be
\lb{g15}
\partial f(m,\b)/\partial m = 0
\ee
If the solution of this equation, $m_{*}(\b)$, appears to be smaller than one,
then for the physical free energy one finds: $f(\b) = f(m_{*}(\b),\b)$.
This situation corresponds to the solid glassy phase, and 
it will be shown to take place only at temperatures
smaller than a certain critical temperature $\T$. At $T \geq \T$
in the high-temperature liquid phase
the formal solution of eq.(\ref{g15}) yields $m_{*}(\b) \geq 1$ 
(which is not allowed by the constraint $m \leq 1$) and 
it can be proved that at the interval $0 \leq m \leq 1$ the maximum
of the free energy $f(m,\b)$ is achieved at $m = 1$. 
Thus, the above scenario looks quite similar to that
of the standard one-step replica symmetry breaking in the random
energy model of spin-glasses \cite{rem}.  

\sectio{The model}

The model we consider in this paper is defined by the following
Hamiltonian:

\be
\lb{g16}
H = -\fr{1}{\sqrt{N}} \sum_{i<j}^{N} U(|{\bf x}_{i}-{\bf x}_{j}|) +
    \fr{1}{2 L} \sum_{i=1}^{N} {\bf x}_{i}^{2} 
\ee
where $\{{\bf x}_{i}\}$ are the coordinates of the point-like particles
in the $D$-dimensional space and

\be
\lb{g17}
U(|{\bf x}|) = \sqrt{2} \cos(|{\bf x}|)
\ee
is the {\it infinite-range} oscillatory interparticle interaction 
potential. The second (Gaussian) part of the Hamiltonian, eq.(\ref{g16}),
bounds the particles to be effectively confined in the finite volume
with the linear size $L$. The parameter $L \gg 1$ is supposed to be
large compared to the oscillation period of the potential, eq.(\ref{g17}),
and it will remain {\it finite} in the thermodynamic limit $N \to \infty$.

The interaction potential, eq.(\ref{g17}), yields
the alternating  concentric bands of positive and negative
energy. In dimensions greater than $D=1$ in a system of $N$ particles 
with the Hamiltonian, eq.(\ref{g16}), it creates
highly complicates "interference" pattern, so that such system appears to be 
strongly "frustrated" in a sense that there exists no optimal position
of the particles which would satisfy all interparticle interactions.
Therefore, a generic solid state (if it exists) in this model must be 
a glass and not the crystal.

The corresponding replica partition function is

\be
\lb{g18}
Z_{m} = \[\prod_{i=1}^{N}\prod_{a=1}^{m} 
        \int \fr{d^{D}{\bf x}_{i}^{a}}{(2\pi L)^{D/2}}\]
        \exp\( \fr{\b}{\sqrt{N}} \sum_{a=1}^{m} \sum_{i<j}^{N}
        U(|{\bf x}_{i}^{a} - {\bf x}_{j}^{a}|) 
        -\fr{\b}{2L} \sum_{a=1}^{m}\sum_{i=1}^{N} ({\bf x}_{i}^{a})^{2}\)
\ee
(for convenience we have introduced here the "natural" normalization factor
$(2\pi L)^{-D/2}$ for the integration over $\{{\bf x}_{i}^{a}\}$).
After the splitting of the degrees of freedom $\{{\bf x}_{i}^{a}\}$ 
into the center of mass $\{{\bf x}_{i}\}$ and the deviations 
$\{{\bf u}_{i}^{a}\}$  (eqs.(\ref{g11})-(\ref{g12})), we get:

\bq
\lb{g19}
 Z_{m} & = & (2\pi L)^{-\fr{NmD}{2}}
        \[\prod_{i=1}^{N}\prod_{a=1}^{m} \int d^{D} {\bf u}_{i}^{a}\]
        \[\prod_{i=1}^{N} m^{D} \d\(\sum_{a=1}^{m} {\bf u}_{i}^{a}\)\]
        \[\prod_{i=1}^{N}\int d^{D} {\bf x}_{i}\] \times
\nn
\nn
&& \times \exp\( \fr{\b}{\sqrt{N}} \sum_{a=1}^{m} \sum_{i<j}^{N} 
      U(|{\bf x}_{i}-{\bf x}_{j}+{\bf u}_{ij}^{a}|) 
      -\fr{\b m}{2L}\sum_{i=1}^{N} {\bf x}_{i}^{2}  
      -\fr{\b}{2L} \sum_{a=1}^{m}\sum_{i=1}^{N}({\bf u}_{i}^{a})^{2}\) 
\eq
where ${\bf u}_{ij}^{a} \equiv {\bf u}_{i}^{a} - {\bf u}_{j}^{a}$.
Keeping only extensive in $N$ terms (the only relevant ones in the 
thermodynamic limit $N \to \infty$) for the (Gaussian) averaging over 
$\{{\bf x}_{i}\}$ we obtain:

\bq
\lb{g20}
&& \[\prod_{i=1}^{N}\int d^{D} {\bf x}_{i} 
     \exp\( -\fr{\b m}{2L} {\bf x}_{i}^{2}\)\]
     \exp\( \fr{\b}{\sqrt{N}} \sum_{a=1}^{m} \sum_{i<j}^{N}
           U(|{\bf x}_{i}-{\bf x}_{j}+{\bf u}_{ij}^{a}|) \) \; =                            
\nn
\nn
&& = \; \(\fr{2\pi L}{\b m}\)^{\fr{ND}{2}}
     \exp\(\fr{\b}{\sqrt{N}} \sum_{a=1}^{m} \sum_{i<j}^{N}
     \ol{U(|{\bf x}_{i}-{\bf x}_{j}+{\bf u}_{ij}^{a}|)} + \right.
\nn
\nn
&& \left.  + \fr{\b^{2}}{2N} \sum_{a,b=1}^{m} \sum_{i<j}^{N}\sum_{k<l}^{N}
     \ol{U(|{\bf x}_{i}-{\bf x}_{j}+{\bf u}_{ij}^{a}|)
         U(|{\bf x}_{k}-{\bf x}_{l}+{\bf u}_{kl}^{b}|)} \)
\eq
where

\be
\lb{g21}
\ol{(...)} \equiv \(\fr{2\pi L}{\b m}\)^{-\fr{ND}{2}}
           \[\prod_{i=1}^{N}\int d^{D} {\bf x}_{i}
           \exp\(-\fr{\b m}{2L} {\bf x}_{i}^{2} \) \] (...) 
\ee

According to the above definition, eq.(\ref{g21}), one can easily prove
(under condition $L \gg 1$):

\be
\lb{g22}
\ol{U(|{\bf x}|)} = 0
\ee

\be
\lb{g23}
\ol{U(|{\bf x_{i}}|) U(|{\bf x_{j}}|)} \simeq \d_{ij}
\ee
and

\be
\lb{g24}
\ol{U(|{\bf x_{i}}|)U(|{\bf x_{j}} + {\bf u}|)} \simeq J_{0}(|{\bf u}|) \d_{ij}
\ee
where $J_{0}(u)$ is the Bessel function. In particular, 
for $u \ll 1$ we have:

\be
\lb{g25}
J_{0}(u\ll 1) \simeq 1 - \2 u^{2}
\ee

Substituting eq.(\ref{g20}) into eq.(\ref{g19}) and 
using eqs.(\ref{g22}) and (\ref{g24}) for the partition function,
eq.(\ref{g19}) we obtain:

\bq
\lb{g26}
 Z_{m} & = & m^{\fr{ND}{2}} \b^{-\fr{ND}{2}}
(2\pi L)^{-\fr{ND(m-1)}{2}}
        \[\prod_{i=1}^{N}\prod_{a=1}^{m} \int d^{D} {\bf u}_{i}^{a}\]
        \[\prod_{i=1}^{N} \d\(\sum_{a=1}^{m} {\bf u}_{i}^{a}\)\]
         \times
\nn
\nn
&& \times \exp\( -\fr{\b}{2L} \sum_{a=1}^{m}\sum_{i=1}^{N}({\bf u}_{i}^{a})^{2}
          + \fr{\b^{2}}{2N} \sum_{a,b=1}^{m} \sum_{i<j}^{N}
 J_{0}(|{\bf u}_{i}^{a} - {\bf u}_{i}^{b} + {\bf u}_{j}^{b} - {\bf u}_{j}^{a}|)\)
\eq
Assuming that all the deviations $\{{\bf u}_{i}^{a}\}$ are small
(only under this condition the separation of the degrees of freedom
defined in eqs.(\ref{g11})-(\ref{g12}) makes physical sense), using
eq.(\ref{g25}), we get

\bq
\lb{g27}
 Z_{m} & = & m^{\fr{ND}{2}} \b^{-\fr{ND}{2}}
(2\pi L)^{-\fr{ND(m-1)}{2}}
        \[\prod_{i=1}^{N}\prod_{a=1}^{m} \int d^{D} {\bf u}_{i}^{a}\]
        \[\prod_{i=1}^{N} \d\(\sum_{a=1}^{m} {\bf u}_{i}^{a}\)\]
         \times
\nn
\nn
&& \times \exp\( \4\b^{2} m^{2} N 
     -\2 (\b^{2} m + \fr{\b}{L}) 
    \sum_{a=1}^{m}\sum_{i=1}^{N}({\bf u}_{i}^{a})^{2} \)
\eq
Simple Gaussian integration over $\{{\bf u}_{i}^{a}\}$
(taking into account the constraints $\sum_{a=1}^{m} {\bf u}_{i}^{a} =0$)
yields:

\be
\lb{g28}
 Z_{m} = \exp\( \4\b^{2} m^{2} N
               -\2 N D m \ln\b 
               -\2 N D (m-1) \ln(1 + \b m L) \)
\ee
For the free energy density, eq.(\ref{g10}), we eventually obtain:

\be
\lb{g29}
f(\b, m) = \fr{D}{2\b} \ln\b -\4 \b m + 
           \fr{D}{2\b} (1 - \fr{1}{m})\ln(1 + \b m L)
\ee

One can easily prove that this function of the parameter $m$ 
has a unique {\it maximum} at $m = m_{*}$ defined by the 
saddle-point equation $\partial f/\partial m = 0$:

\be
\lb{g30}
\fr{1}{2D} (\b m)^{2} = \ln(1 + \b m L) - 
       (1-m)\fr{\b m L}{1 + \b m L}
\ee
At $L \gg 1$ the approximate solution of the above equation
can be obtained explicitly:

\be
\lb{g31}
m_{*}(\b) \simeq \fr{1}{\b} \sqrt{2D \ln L}
\ee
Since the function $f(\b, m)$ is defined only at the interval
$0 \leq m \leq 1$, in the case $m_{*} > 1$ the maximum of $f$ is achieved
at $m=1$. Thus, the maximum of the function $f(\b, m)$ ($0 \leq m \leq 1$)
takes place at

\be
\lb{g32}
m_{*}(\b) = \left\{\ba \fr{1}{\b} \sqrt{2D \ln L} \equiv \fr{\b_{c}}{\b} 
        \; \; ; \mbox{for}  \;  \b > \b_{c}
           \\
           \\
            1 \; \; \; \; \; \; \; \; \; \; \; \; \; \; \; \;  
             \; \; \; \; \; \; \;   ;   \mbox{for} \; \b \leq \b_{c}
           \ea
\right.
\ee
where
\be
\lb{g33}
\b_{c} \simeq \sqrt{2D \ln L}
\ee
Correspondingly, for the physical free energy of the system,
$f(\b) = f(\b, m_{*})$, we obtain the following result:

\be
\lb{g34}
f(\b)  = \left\{\ba 
      \fr{D}{2\b} \ln\b -\4 \b_{c} - 
           \fr{D}{2\b\b_{c}} (\b - \b_{c})\ln(1 + \b_{c} L) 
        \; \; ; \mbox{for}  \;  \b > \b_{c}
           \\
           \\
            \fr{D}{2\b} \ln\b -\4 \b \; \; \; \; \;  \; \; \; \; \; \; \; 
 \; \; \; \; \; \; \; \; \; \; \; \; \; \; \; \; \; \; \; \; \; \; \; \; 
 \; \; \; \; \; \; \;             ;   \mbox{for} \; \b \leq \b_{c}
           \ea
\right.
\ee
One can easily verify that direct (without replicas)
calculation of the free energy

\be
\lb{g35}
f(\b)  = -\fr{1}{\b N} \ln\(\[\prod_{i=1}^{N}\int d^{D} {\bf x}_{i}\]
        \exp\{-\b H[{\bf x}_{i}]\}\)
\ee
yields the same result as in eq.(\ref{g34}) in the high-temperature
region, $\b \leq \b_{c}$. Thus, plain integration over all particles
positions (as it is done in the direct calculations, eq.(\ref{g35}))
correctly describes only the high-temperature (liquid) phase,
but it completely misses the "condensation" of the particles 
into the solid (glassy) phase at $\b > \b_{c}$.

The order parameter which describes the correlations of the particles
inside the replica "molecules" in the glassy phase can be defined 
in the standard way:

\be
\lb{g36}
Q = \la ({\bf u}^{a} - {\bf u}^{b})^{2} \ra
\ee
where $a \not= b$. Since all the replicas are equivalent we can also 
define $Q$ as follows:

\be
\lb{g37}
Q = \fr{1}{m(m-1)} \sum_{a,b}^{m} \la ({\bf u}^{a} - {\bf u}^{b})^{2} \ra
\ee
Using the constraint, eq.(\ref{g12}), we get:

\be
\lb{g38}
Q = \fr{2D m}{(m-1)}  \la (u^{a}_{\a})^{2} \ra
\ee
where the replica number $a$ and the space vector component $\a$
are arbitrary. Note that the above definition of the order parameter is 
valid only for the glassy phase at $m < 1$. In the liquid phase ($m\equiv 1$)
the positions of the particles are not correlated by definition.

Proceeding similarly to the calculation of the replica partition function,
(eq.(\ref{g27})), one finds:

\be
\lb{g39}
\la (u^{a}_{\a})^{2} \ra =
\fr{
\[\prod_{a=1}^{m} \int d^{D} {\bf u}^{a}\] (u^{a}_{\a})^{2}
        \d\(\sum_{a=1}^{m} {\bf u}^{a}\)
        \exp\( -\2 (\b^{2} m + \fr{\b}{L}) 
        \sum_{a=1}^{m}({\bf u}^{a})^{2} \)}{
\[\prod_{a=1}^{m} \int d^{D} {\bf u}^{a}\] 
        \d\(\sum_{a=1}^{m} {\bf u}^{a}\)
        \exp\( -\2 (\b^{2} m + \fr{\b}{L}) 
    \sum_{a=1}^{m}({\bf u}^{a})^{2} \)}
\ee
Straightforward calculations yield:

\be
\lb{g40}
\la (u^{a}_{\a})^{2} \ra = \fr{m-1}{\b m (\b m - L^{-1})}
\ee
Correspondingly for the order parameter, eq.(\ref{g38}), we get:

\be
\lb{g41}
Q = \fr{2D}{\b (\b m - L^{-1})}
\ee
Substituting here the saddle-point value of $m = m_{*}(\b)$ obtained
above, eq.(\ref{g31}), at $L \gg 1$ we eventually find:
 
\be
\lb{g42}
Q \simeq \fr{2D}{\b \sqrt{2D \ln L}} = \fr{2D}{\b\b_{c}}
\ee
We see that in the whole low-temperature region at 
$\b > \b_{c} \simeq \sqrt{2D \ln L} \gg 1$ the value of the 
order parameter $Q$ (as well as the value of the typical 
deviation $\la (u^{a}_{\a})^{2} \ra$) remains small,
which justify the approximation made in calculation of the replica
partition function, eq.(\ref{g27}), as well as the whole original idea
of splitting the degrees of freedom into the centers of masses of the
replica "molecules" and the {\it small} deviations ${\bf u}^{a}$,
eqs.(\ref{g11})-(\ref{g12}).

\sectio{Discussion}

According to eq.(\ref{g36}) the order parameter $Q$ describes the
typical correlations of the particles belonging to the same 
replica molecule. On the other hand, according to the procedure
described in Section 2, the positions of the particles belonging 
to different molecules remain non-correlated, and the value of the 
the fraction of the particles bounded in molecules is controlled
by the parameter $m_{*}(\b)$. According to the standard physical 
interpretation of the replica parameters 
(see e.g. \cite{mpv},\cite{book}), the fraction of particles
localized in space (a kind of a "glass condensate") must be
equal to $(1 - m_{*}(\b)) = 1 - \b_{c}/\b$, and the fraction of
particles which remain free to move must be equal to
$m_{*}(\b) = \b_{c}/\b$. Physically the order parameter $Q$
describes the typical value of the space fluctuations of the 
localized particles around their equilibrium positions.

It is interesting to note that the value of the order parameter
$Q$ computed in the previous Section remains finite in the whole 
low-temperature (glassy) phase including the phase transition point. 
In particular, according to eq.(\ref{g42}) at $\b = \b_{c}$ we have:

\be
\lb{g43}
Q_{c} \simeq \fr{1}{\ln L}
\ee
and $Q(T\to 0) \to 0$. Note also that in the high-temperature 
phase this order parameter is not defined: in terms of replicas
it can be defined only for $m\not= 1$ while at $\b < \b_{c}$,
$m_{*}(\b) \equiv 1$, and in physical terms it describes the 
fluctuation of the localized particles which are just absent
at $\b < \b_{c}$.

Thus, here we face the typical scenario of the phase transition
with one-step replica symmetry breaking well known after the
random energy model of spin glasses \cite{rem}. On one hand, 
in terms of the free energy and others thermodynamical functions 
this is the phase transition of the second order (the free energy
function is continuous at $\T$ and it has a singularity only
in the second derivative over the temperature, see eq.(\ref{g34})),
while on the other hand, the order parameter which characterizes
the low-temperature phase has a finite value right at $\T$
(which corresponds to the first-order phase transitions).

In physical terms the present phase transition can be described 
as follows. Above $\T$ all the particles of the system are 
delocalized and the system is in the liquid phase. Just below $\T$
the glassy phase is characterized by a small fraction 
(equal to $1-T/\T$) of particles which become
localized at random positions in space such that their thermal 
fluctuations around these equilibrium positions described by
the order parameter $Q$ have a finite (small) value
right at $\T$, eq.(\ref{g43}). At lowering the temperature
the fraction of localized particles increases as $(1-T/\T)$
while the value of their thermal fluctuations around 
the localized positions
decrease as $Q(T) \simeq 2D\T T$, eq.(\ref{g42}). Eventually, 
at $T \to 0$ all the particles of the system become localized
in a random positions (without thermal fluctuations)  
such that the system turns into perfectly frozen glass state. 

Of course, from the physical point of view the present model
has two very important defects which makes it not more
than a "toy model" of the glass transition.
First, it has completely non-physical long-range structure of the 
interactions between particles. It this property 
(like in the SK model of spin-glasses) which makes the model 
solvable in terms of the mean-field approach.
Second, the particles in this model are point-like.
Although formally, due to the proper normalization of the 
parameters of the Hamiltonian, the mean-field thermodynamic limit
in this model is quite reasonable (see calculations of Section 3),
nevertheless, from the physical point of view it
looks very strange: here we have an infinite number of particles
which are confined inside a finite volume. Presumably this last drawback
could be cured by introducing a finite size for the particles involved.
On one hand, this should not ruin the mean-field nature of the model,
while on the other hand,
it would make the thermodynamic limit much more reasonable,
where both the number of particles and the volume of the system are
taken infinite while the density of the particles remains finite.

\vspace{15mm}

{\bf Acknowledgments}.

The author is grateful to M.Mezard  and G.Tarjus   
for stimulating discussions.

\end{document}